\documentclass[aps,prb,preprint,unsortedaddress,showpacs]{revtex4}
\usepackage{amssymb}
\usepackage{graphicx}

\begin{document}

\title{Heat capacity of $\alpha$-GaN: Isotope Effects}
\author{R. K. Kremer}
\email[Corresponding author:~E-mail~]{R.Kremer@fkf.mpg.de}
\author{M. Cardona}
\author{E. Schmitt}
\affiliation{Max-Planck-Institut f{\"u}r Festk{\"o}rperforschung,
Heisenbergstr. 1, D-70569 Stuttgart, Germany}

\author{J. Blumm}
\affiliation{ NETZSCH-Ger\"atebau GmbH,
 D-95100 Selb, Germany }
\author{S. K. Estreicher}
\author{M. Sanati}
\affiliation{Physics Department, Texas Tech University, Lubbock,
TX 79409-1051, USA}
\author{ M. Bockowski, I. Grzegory, T. Suski}
\affiliation{Institute of High Pressure Physics, Polish Academy of
Sciences, ul Sokolowska, 01-142 Warszawa, Poland}

\author{A. Jezowski}

\affiliation{ Institute of Low Temperature and Structure Research,
Polish Academy of Sciences, ul. Okolna, 50-422 Wroclaw, Poland }
\date{\today}

\begin{abstract}
Until recently, the heat capacity of GaN had only been measured
for polycrystalline powder samples. Semiempirical as well as
\textit{first-principles} calculations  have appeared within the
past few years. We present in this article measurements of the
heat capacity  of hexagonal single crystals of GaN in the 20-1400K
temperature range. We find that our data deviate significantly
from the literature values for polycrystalline materials. The
dependence of the heat capacity on the isotopic mass has also been
investigated recently for monatomic crystals such as diamond,
silicon, and germanium. Multi-atomic crystals are expected to
exhibit a different dependence of these heat capacities on the
masses of each of the isotopes present. These effects have not
been investigated in the past. We also present
\textit{first-principles} calculations of the dependence of the
heat capacities of GaN, as a canonical binary material, on each of
the Ga and N masses. We  show that they are indeed different, as
expected from the fact that the Ga mass affects mainly the
acoustic, that of N the optic phonons. It is hoped that these
calculations will encourage experimental measurements of the
dependence of the heat capacity on isotopic masses in binary and
more complex semiconductors.
\end{abstract}

\pacs{65.40.+g; 63.20.Dj} \maketitle


 \email{rekre@fkf.mpg.de}

\section{Introduction}
The investigation of the heat capacity of crystals is an old topic
of condensed matter physics with which illustrious names were
early associated \cite{Einstein,Nernst,Debye}. Knowledge  of the
heat capacity of a substance not only  provides  essential insight
into its vibrational properties but is also mandatory for many
applications.

Two famous limiting cases are correctly predicted by the standard
elastic continuum theory \cite{Debye}: At high temperatures, the
constant-volume heat capacity, $C_v$, tends to the Petit and
Dulong limit, e. g. $\sim$ 49.9 J/mol K in binary materials such
as GaN\cite{Dulong}.  At sufficiently low temperatures, $C_v$ is
proportional to $T^3$ (Ref. (\onlinecite{Debye})). At intermediate
temperatures, however, the temperature dependence of $C_v$ is
governed by the details of vibrations of the atoms and for a long
time could only be determined from experiments. In recent years,
the temperature-dependence of the specific heat of many systems
has been calculated in the harmonic approximation. The
calculations ranged from semiempirical, based on inelastic neutron
scattering or X-ray data, to \textit{first-principles}, based on
total energy calculations.

For example, for GaN,  a very topical semiconductor these days,
the temperature dependence of the specific heat capacity  has been
calculated, in the harmonic approximation, on the basis of either
experimental or based on \textit{first-principles} phonon
frequencies. The phonon frequencies were obtained semiempirically,
from inelastic neutron \cite{Nipko} or x-ray scattering data
\cite{Serrano} or from \textit{first-principles}, based on
electronic calculations of total energies \cite{Sanati}.

In the past 15 years a variety of semiconductor crystals with
different isotopic compositions have been grown \cite{Itoh}. Using
these crystals, the effects of the isotopic mass on the phonon
spectra \cite{Widulle}, in particular the related anharmonic
effects, have been investigated. Of considerable interest has also
been the dependence of the electronic spectra on isotopic mass,
which is effected by the electron-phonon interaction
\cite{Cardona2004}. The influence of the isotopic masses on
thermo-mechanical properties, including the thermal conductivity
\cite{Asen,Kremer,Ruf,Wei} the thermal expansion \cite{Pavone}
 and the specific heat
\cite{CardonaCP,Gibin,SanatiCard} have also been investigated.
Monatomic materials such as diamond, Si, and Ge have only one
average isotopic mass that can be varied (effects of fluctuations
in the isotopic mass will not be discussed here). GaP and GaAs are
binary compounds but their anions have only one stable isotope,
hence only the effects of varying the cation isotope mass can be
investigated (even so, very interesting effects have been seen in
the Raman phonons of GaP (see Ref. \onlinecite{WidullePRL})).

GaN is an ideal material to investigate isotope effects: Ga
possesses two stable isotopes, ($^{69}$Ga and $^{71}$Ga), whereas
N has also two such isotopes, $^{14}$N and $^{15}$N. An
investigation of the electronic energy gap of GaN and its
dependence on the isotopic mass of N was published recently
\cite{Manjon}. Corresponding measurements in which the isotopic
mass of gallium is varied were, at that time, not possible because
of the lack of suitable isotopically modified samples: they are
being performed now.

Isotopically modified samples of GaN large enough to measure the
dependence of the heat capacity on the masses of Ga and N are
still not available. In this paper we present first-principles
results concerning this problem, in the hope that they will
encourage crystal growth with isotopically pure Ga and N. Our
theoretical results are displayed, in the Debye manner, as $C_v
/T^3$, for the four possible isotope combinations, compared with
the calculations for GaN with the natural isotopic abundances. The
maximum in $C_v /T^3$ found at about 40 K increases when any of
the isotopic masses is increased, this effect for gallium being
larger than for nitrogen. The effect of the latter becomes larger
than that of the former at higher temperatures. We also present
the calculated logarithmic derivative of  $C_v /T^3$ vs. each of
the two masses: the effect of the mass of nitrogen is shown to be
displaced to higher temperatures with respect to that of the mass
of gallium. This can be qualitatively attributed to the fact that
the gallium mass affects mainly the acoustic phonons of GaN
whereas that of N affects mainly the optical phonons. The
temperature dependence of the logarithmic derivatives of  $C_v
/T^3$ is compared with the results obtained for monatomic
crystals, in which only one mass can be varied.

Due to the difficulties involved in growing large enough single
crystals, the experimental results available thus far for GaN have
been obtained for polycrystalline powder samples only
\cite{Koshchenko,Leitner,Yamaguchi,Chen}. In this article, we
present new measurements of the heat capacity
 in the range 20-1400 K of single-crystals of GaN grown by the
 high-pressure, high-temperature technique.
Our new heat capacity data are in good agreement with the results
of the \textit{first-principles} calculations. However,
particularly above room temperature, the heat capacities
previously obtained on powder samples, and commonly quoted in the
literature as standard values, deviate significantly from our data
and \textit{first-principles} theory. We therefore suggest that
the standard values for the heat capacity of GaN should be
reconsidered and replaced.

\section{Experimental}
GaN single crystals were grown by a self-seeding process at
pressures of nitrogen of about 1.5 GPa and temperatures of 1800 K
from a nitrogen solution in a droplet of gallium
\cite{Karpinski,Grzegory1998,Grzegory2001}. The crystals used in
this experiment were of  up to 5 mm$^2$ lateral size and thickness
of 40-300 $\mu$. They have a good crystallographic quality (x-ray
diffraction rocking curves of 30-100 arcsec). They were highly
conductive (n $>$ 5$\times$10$^{19}$ cm$^{ -3}$) due to
unintentionally introduced oxygen donors but showed a very low
concentration of dislocation which did not exceed 10$^2$
cm$^{-2}$. After removing the crystals  from the high pressure
growth chamber they were etched in aqua regia which can result in
a traces of Ga$_2$O$_3$ on the etched surfaces of crystal plates.

The heat capacities in the temperature range 15 K $<T<$ 400 K were
measured using a commercial Physical Property Measurements System
calorimeter (Quantum Design, 6325 Lusk Boulevard, San Diego, CA.)
employing the relaxation method. To thermally anchor the crystals
to the sample holder platform, a minute amount of Apiezon N grease
was used. The total heat capacity of the platform and grease was
determined in a separate run and subtracted from the measurements
for the GaN crystals.  The heat capacities in the temperature
range 300 K $<T<$ 750 K were measured with a Perkin-Elmer Pyris 1
Differential Scanning Calorimeter (Perkin Elemer, 45 William
Street, Wellesley, MA 02481-4078 U.S.A.) in a Pt crucible. A total
of three independent runs were merged and fitted with a
polynomial. The deviation of the experimental points from this
polynomial was always less than $\pm$1.5\%. The maximum observed
difference between  data  points for the different runs was
$\simeq$2\%.
 In the temperature range 350 K $<T<$ 1400 K we used a
Netzsch 404C Pegasus DSC calorimeter  (NETZSCH-Ger\"atebau GmbH,
Wittelsbacher Str. 42, D-95100 Selb, Germany) with the sample
contained in a Pt crucible equipped with alumina liners. For the
DSC measurements several crystals were combined to obtain samples
with masses between 20 and 60 mg. During the measurement the DSC
calorimeters were flushed with N$_2$ gas.  The calibration of the
DSC calorimeters was done with a sapphire crystal.

\section{First-Principles Calculations}

The theoretical results described below were obtained from
self-consistent, \textit{first-principles} calculations based on
local density-functional theory with the SIESTA
code\cite{siesta1,siesta2}, as described in Ref.
\onlinecite{SKEthermo}. The exchange-correlation potential of
Ceperley-Alder\cite{ca}, parameterized by Perdew and
Zunger\cite{pz}, was used. Norm-conserving pseudopotentials in the
Kleinman-Bylander form\cite{kb} were used to remove the core
regions from the calculations. The basis sets for the valence
states were linear combinations of numerical atomic orbitals
(double-zeta with a set of $d$ orbitals for Ga). The host crystal
was represented by a (wurtzite) Ga$_{36}$N$_{36}$ periodic
supercell. The calculated lattice constants and bulk modulus were
found to be $a$ = 3.200 \AA, $c/a$ = 1.629, and $B$ = 211 GPa,
respectively, which compare well with the literature
values\cite{Tsuchiya} (3.189 \AA, 1.607, and 202 GPa,
respectively).

\begin{figure}[tbph]
\includegraphics[width=12cm ]{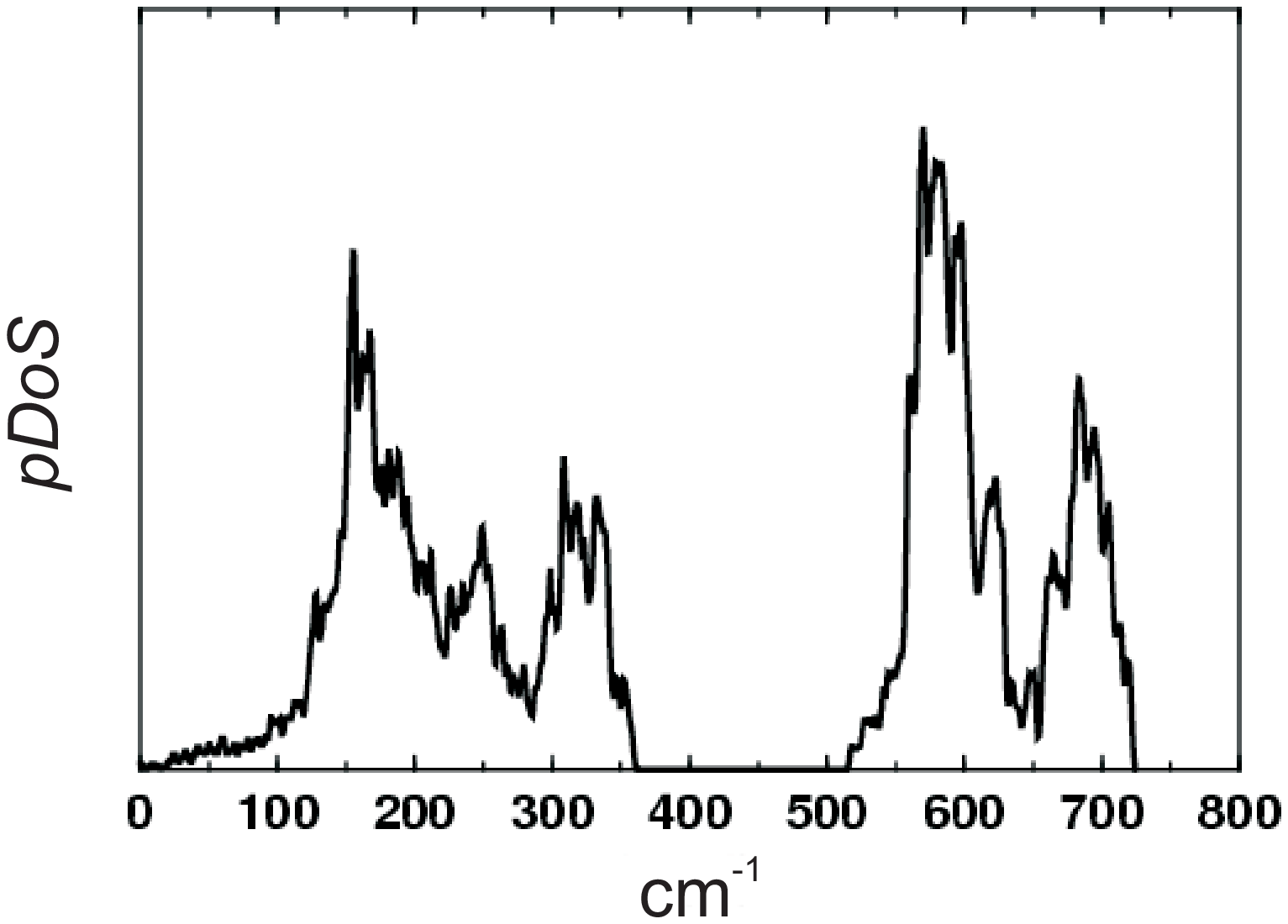}
\caption{Calculated phonon density of states (pDoS) of GaN taken
from Ref. \onlinecite{Sanati}. The band below 400 K corresponds
basically to Ga vibrations (except in the acoustic region below
$\approx$ 100 K where both atoms vibrate). The optical band, above
500 K, corresponds to nitrogen vibrations.} \label{pdos}
\end{figure}

The matrix elements of the (harmonic) dynamical matrices were
extracted at $T$ = 0 K from the derivatives of the density matrix
relative to nuclear coordinates using the perturbative approach
developed by Baroni {\it et al.}\cite{baroni} and Gonze {\it et
al.}\cite{gonze} and implemented into SIESTA by Pruneda {\it et
al.}\cite{jmap}. The dynamical matrix  was then evaluated at 120
$q$ points along high-symmetry directions of the  Brillouin zone.
This procedure generated some 2.6$\times$10$^4$ normal mode
frequencies. The phonon density of states (pDoS) shown in Fig.
\ref{pdos} was obtained by smoothing the normal mode frequencies
with Gaussians with a width of 1 cm$^{-1}$.

In the harmonic approximation, the Helmholtz free energy is given
by\cite{maradudin}

\begin{equation}
F_{vib.}(T)=k_BT\int_0^\infty\>\ln\{\sinh (\hbar\omega/2k_BT)\}
g(\omega)d\omega
\end{equation}

\noindent where $k_B$ is the Boltzmann constant. Here, the pDoS
$g(\omega)$ is normalized so that $\int g(\omega)d\omega=3N$,
where $N$ is the number of atoms. Note that $F_{vib.}(T=0)$ is the
total zero-point energy. Once $F_{vib.}$ is calculated, the
vibrational entropy and specific heat at constant volume are given
by

\begin{equation}
S_{vib.}=-\left(\frac{\partial F_{vib.}}{\partial
T}\right)_v\qquad C_v
          =-T\left(\frac{\partial^2 F_{vib.}}{\partial T^2}\right)_v\>.
\end{equation}

\begin{figure}[tbph]
\includegraphics[width=12cm ]{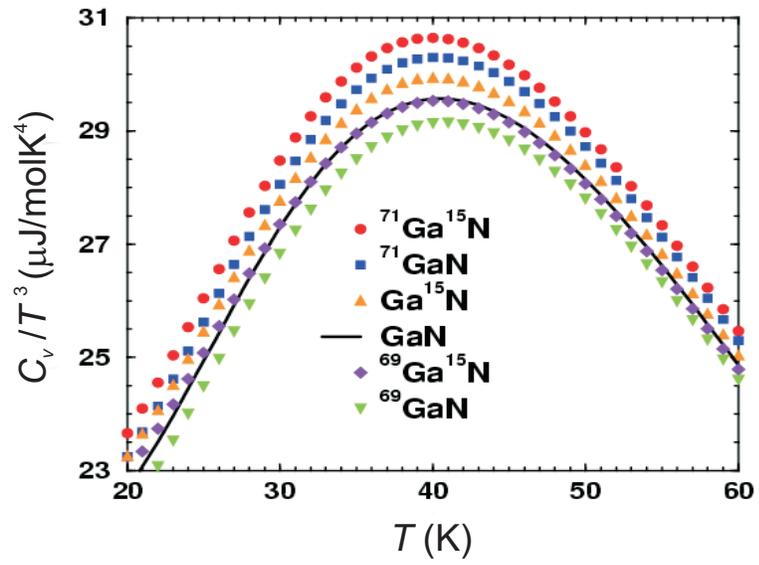}
\caption{(color online) Calculated $C_v/T^3$ for GaN with various
isotope compositions. } \label{CT3thiso}
\end{figure}

\begin{figure}[tbph]
\includegraphics[width=12cm]{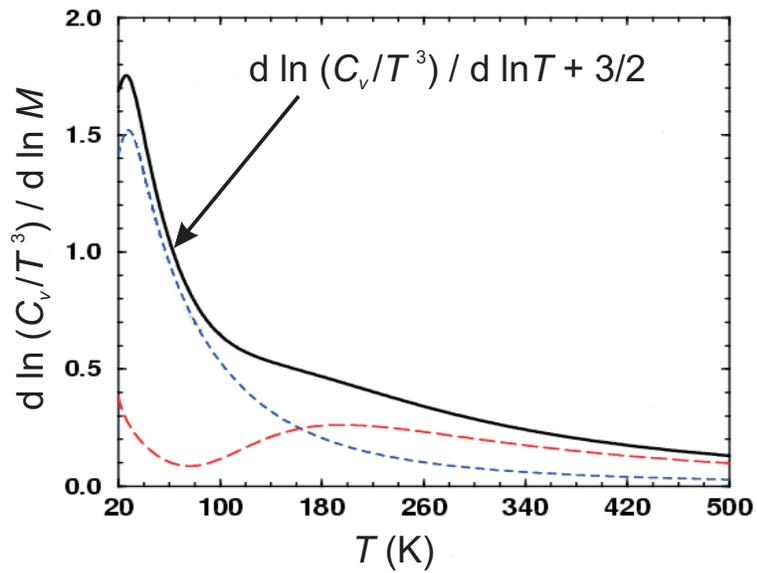}
\caption{(color online) Calculated logarithmic derivatives $d\ln
(C_v/T^3)/d\ln T +3/2$ (right-hand- side of Eq. (\ref{eq1}) (solid
black line) compared to the left-hand-side $d\ln (C_v/T^3)/d\ln M$
with $M=M_{Ga}$ (short-dash line, blue) and $M=M_N$ (long-dash
line, red). Note that the sum of the two dashed lines is roughly
equal to the solid black line.
 } \label{dCtodT}
\end{figure}

In Fig. \ref{CT3thiso}, we display $C_v/T^3$  calculated in the
20-60 K range for six isotopic compositions, including natural Ga
( natural nitrogen can be assumed to be pure $^{14}$N for our
purposes). This figure clearly illustrates  that the effect of
changing the mass of Ga is considerably larger than that of
changing the mass of N in the
 temperature region under consideration. This is because the vibrations of
nitrogen take place at much higher frequency than those of Ga (see
Fig.\ref{pdos}).  Figure \ref{dCtodT} shows the logarithmic
derivatives of $C_v/T^3$ vs. the masses of either the Ga or the N
atoms, as well as vs. temperature calculated from the data of Fig.
\ref{CT3thiso}. This plot was particularly useful for monatomic
crystals because it could be related to either the measured
\cite{Gibin,CardonaCP} or calculated temperature dependence of
specific heat by the equation\cite{SanatiCard}:

\begin{equation}
\frac{d \ln (C_v/T^3)}{d \ln M} =  \frac{d \ln (C_v/T^3)}{d \ln T}
+ \frac{3}{2}. \label{eq1}
\end{equation}

We have also plotted in Fig. \ref{dCtodT}  a curve obtained with
Eq. \ref{eq1} using the data of Fig. \ref{CT3exp}. In comparing
this plot with the calculated 0 K values of both derivatives, we
realize that the results of Eq. (\ref{eq1}) correspond roughly to
the sum of the two separate derivatives. This is understood by
considering that in the monatomic case the masses of both atoms in
the unit cell are varied when taking the mass derivative, whereas
in the binary case we vary only one at a time. In the limit ($T
\rightarrow 0$ ) Eq. (\ref{eq1}) yields a derivative equal to 3/2.
The logarithmic derivatives vs. mass in the binary case (e.g. GaN)
are then  given by:

\begin{equation}
\frac{d \ln C_v}{d \ln M_{\rm Ga}} =  \frac{3}{2} \frac{M_{\rm
Ga}}{M_{\rm Ga} + M_{\rm N}} = 1.25,  \label{eq2a}
\end{equation}

\begin{equation}
\frac{d \ln C_v}{d \ln M_{\rm N}} = \frac{3}{2} \frac{M_{\rm
N}}{M_{\rm Ga} + M_{\rm N}} = 0.25. \label{eq2b}
\end{equation}

 The sum of both derivatives in Eqs. (\ref{eq2a},\ref{eq2b}) equals 3/2, as predicted
 for a monatomic crystal.
This trend can already be seen in the derivatives shown in Fig.
\ref{dCtodT} for 20 K. In this figure, the gallium derivative
remains larger than the nitrogen derivative up to about 160 K. At
this temperature the contribution of the gallium to $C_v$  has
already flattened out considerably, while tending to the Petit and
Dulong limit, whereas that of nitrogen has begun to rise up and
takes over.

Notice that below 80 K the derivative vs. $M_{\rm N}$ increases
down to 20 K, the lowest temperature we could reliably reach in
our calculations. This can be attributed to a dominant
contribution of long-wavelength acoustic phonons which include
vibrations of Ga as well as of N (the appropriate effective mass
is $M_{\rm Ga}$ + $M_{\rm N}$). The maximum in the $M_{\rm N}$
derivative seen in Fig. \ref{dCtodT} at $\sim$ 220 K corresponds
to that found for the $M_{\rm Ga}$ derivative at 30 K. It signals
the rising contribution of the nitrogen related acoustic phonons.

\section{Experimental Results and Discussion}

\begin{figure}[tbph]
\includegraphics[width=12cm ]{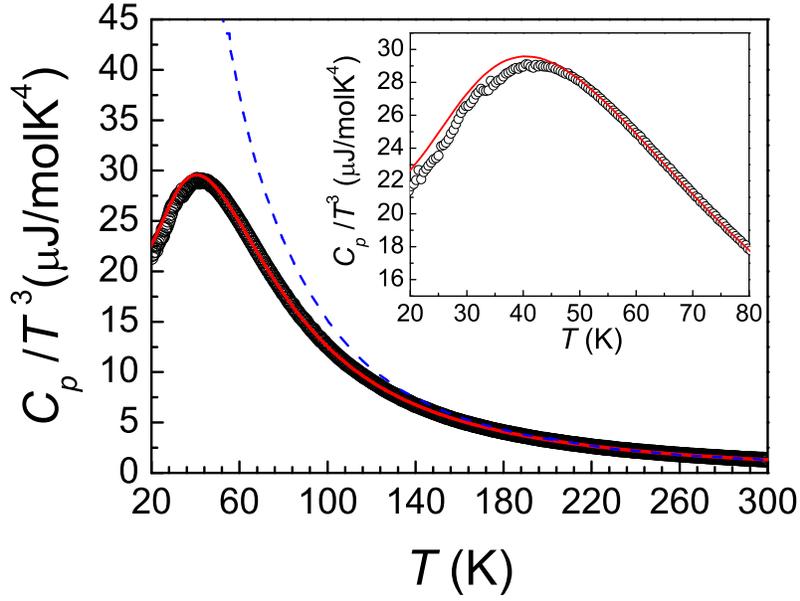}
\caption{(color online) Low temperature experimental molar heat
capacity $C_p/T^3$ of wurtzite GaN
     (circles). The (red) solid line represents the results of the
\textit{first-principles} calculations for $C_v / T^3$ ($C_v
\approx C_p$). The (blue) dashed line represents the  data
obtained by  Koshchenko \textit{et al.} on a polycrystalline
sample \cite{Koshchenko}. The inset shows the region of the Debye
maximum in an enlarged scale.} \label{CT3exp}
\end{figure}

Figure  \ref{CT3exp} displays our low temperature molar heat
capacity data of crystalline GaN as a Debye plot ($C_p / T^3$ vs.
$T$), together with the results of the \textit{first-principles}
calculations in the temperature range 20 K $< T< $ 300K. For these
temperatures,  thermal expansion contributions to the heat
capacity are negligible, and to a very good approximation, $C_p
\approx C_v$ (cf. the discussion below).

The agreement of our results with the \textit{first-principles}
calculations is very good, small deviations ($<$ 2.5\%) are
visible below the maximum at $\sim$ 40 K. The magnitude of the
Debye maximum is well reproduced in the experimental data, $T_{\rm
max}$ is shifted by 1.5 K to higher temperatures from the value
predicted by the calculations. For comparison we also show the
data by Koshchenko \textit{et al.} \cite{Koshchenko} obtained from
measurements of a polycrystalline sample. Below $\sim$ 200 K
Koshchenko's data, taken so far as the standard values for the low
temperature heat capacity of GaN, deviate significantly from the
\textit{first-principles} results and the present experimental
data. They completely fail to reproduce the maximum at $\sim$ 40
K, which is due to deviations from a $T^3$ power law behavior at
low temperatures. We presume that impurities may be the reason for
this failure. Above 150 K, Koshchenko's results agree fairly well
with our data and the \textit{first-principles} results. The heat
capacities determined by Leitner \textit{et al.} for temperatures
200 K $< T < $ 1300 K (not shown in Fig. \ref{CT3exp})
consistently lie above our data. Up to 300 K the difference does
not exceed 2\%.

We discuss next our heat capacity results in the the temperature
range 300 K $ < T < $ 1400 K where the difference between
constant-volume and constant-pressure specific heats becomes
increasingly important. The constant-volume and constant-pressure
specific heats are related by \cite{Ashcroft}

\begin{equation}
C_p(T) = C_v(T) + E(T) \cdot T,
 \label{Ashcroft}
\end{equation}

\noindent with

\begin{equation}
E (T) =  \alpha_v^2(T) \cdot B \cdot V_{\rm{mol}},
 \label{expansion}
\end{equation}

\noindent
 where $\alpha_v(T)$ is the temperature dependent volume coefficient of thermal
expansion, $B$ the bulk modulus and $V_{\rm mol}$ the molar volume
at $T$ = 0 K. The linear thermal expansion coefficients,
$\alpha_a(T)$ and $\alpha_c (T)$, have been measured by x-ray
diffraction over a wide temperature range (100 K $< T < $ 1500 K)
by Wang and Reeber \cite{Wang}. The volume thermal expansion
coefficient, $\alpha_v (T)$ = 2 $\alpha_a (T)$+$\alpha_c(T)$,
amounts to 12.29$\times$10$^{-6}$ K$^{-1}$ at 300 K and it
increases to 15.57$\times$10$^{-6}$ K$^{-1}$ at 1500 K. Above 200
K the overall temperature dependence of the volume thermal
expansion coefficient can be well described by $\alpha_v(T$) =
15.59(4)$\times$10$^{-6}$ K$^{-1}\cdot$(1 -
0.25(1)$\times$10$^5$K$^2$/$T^2$ +
0.50(8)$\times$10$^9$K$^4$/$T^4$). The overall agreement of this
fit  with the experimental results is better than $\pm$1\%.

\begin{figure}[tbph]
\includegraphics[width=14cm ]{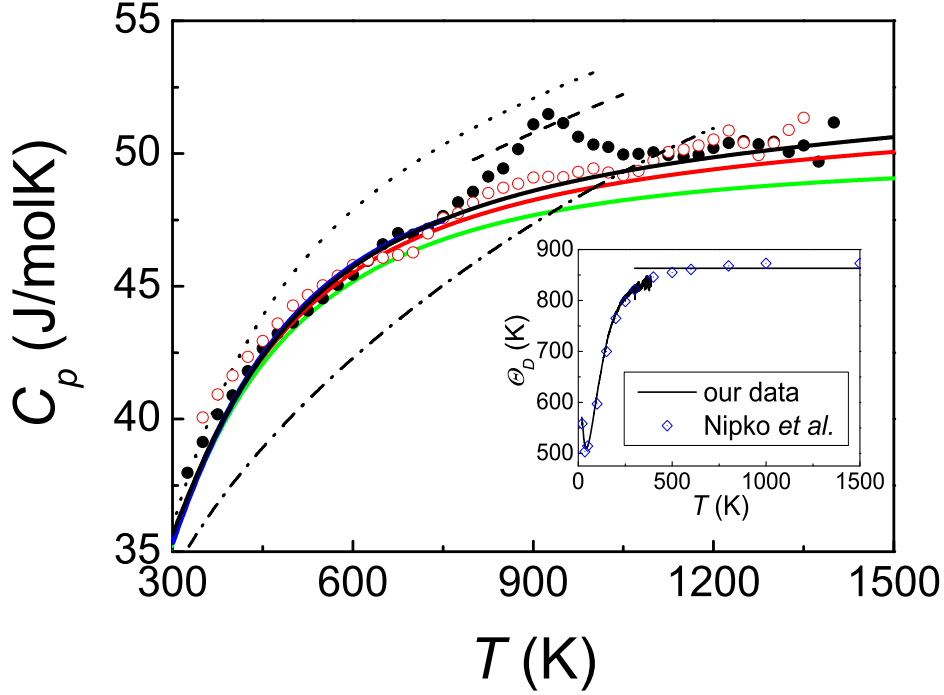}
\caption{(color online) The experimental heat capacities $C_p$ of
wurtzite GaN crystals compared with theoretical results. The solid
(black) line represents the best fit of Eq. (\ref{cpdebye}) to our
data combining all our data (except the anomaly around 600 K, see
text). The solid (green) line represents the
\textit{first-principles} $C_v$ data by Sanati \textit{et al.}
(Ref. \onlinecite{Sanati}). The solid (red)  line shows $C_p$ as
calculated according to Eqs. (\ref{Ashcroft},\ref{expansion})
using the literature values for the bulk modulus and the molar
volume and  the expansion coefficients for the volume thermal
expansion, $\alpha_v (T)$, as given in the text. The high
temperature data collected with the Netzsch calorimeter
($\bullet$, $\circ$) show an anomaly at $\sim$ 900 K, most likely
due to the sublimation of traces of Ga$_2$O (see text). A
concomitant small weight loss has been observed in a TGA trace in
the respective temperature region. The dotted (black), the dashed
(black) and the dash-dotted lines show literature data by Leitner
\textit{et al.} (Ref. \onlinecite{Leitner}), Itagaki \textit{et
al.} (Ref. \onlinecite{Yamaguchi}) and Chen \textit{et al.} (Ref.
\onlinecite{Chen}), respectively. The inset displays the Debye
temperatures corresponding to our data according to the Debye law
(20 K $ < T < $ 400 K), in comparison with Nipko's results
\cite{Nipko}. To fit the high temperature data with Eq.
(\ref{cpdebye}) a temperature-independent Debye temperature
$\Theta_{\infty}$, indicated by the horizontal bar, has been
assumed.}
 \label{Allheat}
\end{figure}

 Figure \ref{Allheat} shows our measured molar heat
capacities, $C_p$,  of crystalline GaN in comparison  with the
data available in the literature for polycrystalline GaN. The data
by Leitner \textit{et al.} \cite{Leitner} and Yamaguchi \textit{et
al.} \cite{Yamaguchi}, obtained on powder samples and commonly
quoted as standard values for the specific heat of GaN at high
temperatures,  and in particular the data by Chen \textit{et
al.}\cite{Chen}, deviate considerably  from our results. Above 800
K,  our data taken on a pristine sample, reveal a broad hump
centered at about 900 K. This anomaly is no longer seen  in a
subsequent second run carried out on the same sample (see the
discussion below). Above 1100 K, the results of both runs agree
again within experimental error.

The agreement of our high temperature heat capacity data with $C_p
= C_v + E(T)\cdot T$, with $C_v$ obtained from \textit{first
principles} by Sanati and Estreicher  \cite{Sanati}, is within the
expectations for such calculations. To calculate the thermal
expansion contribution $E(T)$ according to Eq. (\ref{expansion}),
the bulk modulus $B$ = 202.4 GPa (Ref. \onlinecite{Tsuchiya}), the
molar volume $V_{mol}$ = 13.76$\times$10$^{-6}$ m$^3$ and the
polynomial fit of the temperature dependent $\alpha_v (T)$
described in detail above were used.

The solid (black) line in Figure \ref{Allheat} was obtained by
least-squares fitting all our data (for $T \gtrsim$ 300 K,
excluding the anomaly around 900 K seen in the first run)
simultaneously to a power series expansion of the Debye integral,
including terms $\propto 1/T^2$ and $\propto 1/T^4$ as well as a
term  $C\cdot T$ to account for the thermal expansion. For
simplicity we have assumed a temperature independent Debye
temperature $\Theta_\infty $ (Ref. \onlinecite{Tolman}).

\begin{equation}
C_p(T) = 2\times 3\,R\,\, (1 -
\frac{1}{20}\,\frac{\Theta_\infty^2}{T^2} +
\frac{1}{560}\,\frac{\Theta_\infty^4}{T^4}) + C\cdot T,
 \label{cpdebye}
\end{equation}

\noindent with $R$ being the gas constant.

Our measured molar heat capacity $C_p$ in the temperature range
300 K $< T <$ 1400 K can be best described with $\Theta_\infty$ =
863(3) K. This Debye temperature correlates very well with the
$\Theta_D (T)$ obtained from the measurements below room
temperature and is consistent with  $\Theta_D (T)$ calculated by
Nipko \textit{et al.}\cite{Nipko} from inelastic neutron
scattering data. (see inset in Fig. \ref{Allheat}).

The fit with Eq.(\ref{cpdebye}) gives $C$ = 1.03$\pm$0.01 mJ/mol
K$^2$, somewhat larger than the  value of $\sim$ 0.7 mJ/mol K$^2$
expected using the thermal expansion coefficient for $T
\rightarrow \infty$ given above.

There is an  anomaly centered at $\sim$ 900 K, but starting
already at $\sim$ 700 K, in the heat capacity data collected in
the first run up to 1400 K. This feature was no longer seen in a
subsequent run carried out on the identical sample. Additionally,
we detected a small weight loss starting at $\sim$ 600 K which
finally amounts to $\sim$ 0.05\% at 1400 K. The heat capacity
anomaly and the weight loss indicate  sublimation, most likely of
the Ga-suboxide species, Ga$_2$O. Sublimation of Ga$_2$O in
high-vacuum has been found to commence  at 500 $^{\rm o}$C
\cite{Brukl,Klemm}. Since the DSC measurements were carried out in
N$_2$ atmosphere Ga$_2$O could  result from a reduction of
adherent traces of Ga$_2$O$_3$ detected after the freshly grown
crystals have been etched with aqua regia (see above).

In summary, we determined the heat capacity of GaN single crystals
in the temperature range 20 K $ < T <$ 1400 K and found
significant deviations from the values given in the literature for
polycrystalline material. Good agreement of our experimental data
with \textit{first principles} calculations is observed over the
full temperature range. We make detailed predictions as to how the
heat capacity will vary if the isotope masses of Ga and N are
changed. We hope that these interesting results will soon be
confirmed by measurements on isotopically pure GaN crystals.

\begin{acknowledgments}
We thank G. Siegle for expert experimental assistance and U. Klock
for preliminary high temperature heat capacity measurements (not
reported here). The works of SKE is supported in part by a grant
from the R.A. Welch Foundation.

\end{acknowledgments}

\end{document}